\documentclass{PoS}

\usepackage{graphicx}
\usepackage{multicol}

\newcommand{\treeC}[1] {\left[#1\right]_{C_{i}}} 
\newcommand{\be}{\begin{equation}}
\newcommand{\ee}{\end{equation}}
\newcommand{\ba}{\begin{eqnarray}}
\newcommand{\ea}{\end{eqnarray}}

\title{Determination of Low Energy Constants and testing Chiral Perturbation
  Theory at Next to Next to Leading Order}

\ShortTitle{Determination of LECs and testing ChPT at NNLO}

\author{Johan Bijnens\\
        Department of Theoretical Physics, Lund University,
        S\"olvagatan 14A, SE 22362 Lund, Sweden\\
        E-mail: \email{bijnens@thep.lu.se}}

\author{\speaker{Ilaria Jemos}\\
        Department of Theoretical Physics, Lund University,
        S\"olvagatan 14A, SE 22362 Lund, Sweden\\
        E-mail: \email{ilaria.jemos@thep.lu.se}}

\abstract{We present the results of a search for relations between
observables that are independent of the Chiral Pertubation Theory (ChPT)
Next-to-Next-to-Leading Order (NNLO) Low-Energy Constants (LECs). We have found
some relations between observables in $\pi\pi$, $\pi K$ scattering and
$K_{l4}$ decay which have been evaluated numerically using the old fit (fit 10
in~\cite{Amoros:2001cp}) of the
NLO LECs.}

\FullConference{International Workshop on Effective Field Theories: 
                from the pion to the upsilon \\
                February 2-6 2009\\
                Valencia, Spain}

\renewcommand{\logo}
{\raisebox{-5mm}
{\parbox{\textwidth}{\begin{flushright}
LU TP 09-10\\
arXiv:yymm.nnnn [hep-ph]\\
April 2009
\end{flushright}
\includegraphics{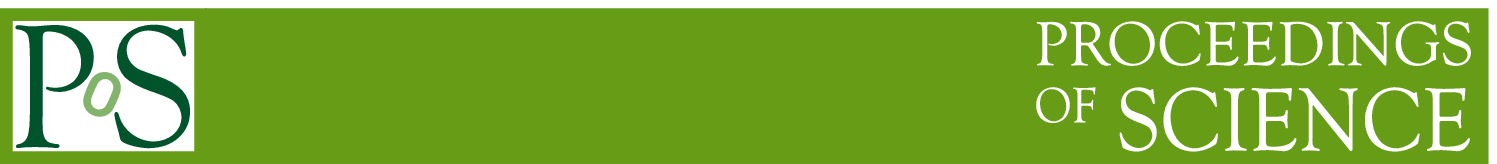}}
}}
\FullConference{{\rm To be published in the proceedings of:\\}
International Workshop on Effective Field Theories: 
  from the pion to the upsilon\\
                 2-6 February 2009\\
                 Valencia, Spain}

\begin{document}

\section{Introduction}

As pointed out in many talks in this conference one of the major problems of
ChPT is the determination of the LECs. This is an issue that has to be solved if
we want a complete predictive theory and to check its
convergence. Furthermore, since the LECs encode the dynamics of the underlying
theory QCD, they can in principle provide us with more information on it as
well.

On the other hand, chiral symmetry imposes no constraints on the values of the
coupling constants, thus we need to perform a fit for their determination. This
is a rather difficult task for different reasons.

First of all, going to higher order in the chiral
expansion the number of independent operators allowed by the symmetries
in the Lagrangians, and
therefore the number of coupling constants to estimate, increases. E.g.,
in $SU(3)$ ChPT up to NNLO, the following LECs appear:
$2$ in $\mathcal{L}_2$ $(F_0,B_0)$, $10+2$ in $\mathcal{L}_4$ $(L_is, H_is)$
and $90+4$ in $\mathcal{L}_6$ $(C_i)$s.

Moreover these
constants are strongly correlated. As a matter of fact since at
order $p^6$ typically many $L_i$s
contribute to particular processes, their determination entangles different
processes. As a result, an estimate of an order $p^6$ LEC used in one process
where an $L_i$ is determined sneaks in the determination of the other $L_i$s
and possibly of the $C_i$s in the other processes.
The solution, a full comprehensive analysis of all processes at the same time,
is a major undertaking which has not been done~\cite{Bijnens:2006zp}.

Finally, so far we don't have enough data to perform a complete fit of
all the constants, even if in this regard 
other kinds of results, such as dispersive and lattice calculations, are
helpful.

The solution which has been mainly used so far is to perform the fit of the
$L_i$s relying on estimates of the values of the $C_i$s by simple resonance
saturation, see the discussion in \cite{Amoros:2001cp,Amoros:2000mc}, but
now we have at our disposal a lot of processes calculated up to NNLO
(see~\cite{Bijnens:2006zp} for a review) and new measurements of the
obsevables involved, thus it is
time to collect all this knowledge and perform a new global fit.

As said above, one of the main problems to overcome when performing the
fit is the large number
of unknown constants appearing at NNLO. For this purpose we have looked for
relations between observables that do not involve the $C_i$.

If $O$ is an observable, then ChPT allows us to write it is as a sum of terms of increasing importance in the chiral expansion:
\ba
O&=&O^{(2)}+O^{(4)}+O^{(6)}\,.
\ea
The $p^{6}$ part can be split as
\ba
O^{(6)}&=&O_{C_{i}{\rm(tree\,level)}}+O_{L_{i}\rm{(one\,loop)}}
+O_{F_{0}\rm{(two\,loops)}}.
\ea
We found relations between observables such that the first
contribution, the only one where the $C_i$ dependence shows up,
cancels out. Using these relations we can stop worrying about the $C_{i}$s and
perform the fit of the $L_{i}$s at NNLO.\footnote{However often
the tree level contribution from the $L_i s$ also cancels.} 
Moreover we can check how large
the loop contributions are and thus test ChPT convergence.

So far we considered the following processes and quantities: $\pi\pi$ and
$\pi K$ scattering, $K_{l4}$ $(K\rightarrow\pi\pi e\nu)$, scalar form
factors $(F_{S}^{\pi /K}(t))$, meson masses, meson decay constants
($F_{\pi/K}$), vector form factors $(F^{\pi / K}_{V})$ and
$\eta\rightarrow\pi\pi\pi$.
We found many relations, but not all of them are equally useful for the
fit purpose: some of them involve not yet well known observables
and some others are
long and complicated expressions. Hence in the following we only quote
the most relevant ones. All results presented are preliminary.
We discuss now the relations and then a first numerical check of
some of them.

\section{Relations between Observables}

\subsection{$\pi\pi$ scattering}

The $\pi\pi$ scattering amplitude can be written as a function $A(s,t,u)$
which is symmetric in the last two arguments:
\begin{equation}
A(\pi^{a}\pi^{b}\rightarrow\pi^{c}\pi^{d})=
\delta^{a,b}\delta^{c,d}A(s,t,u)+\delta^{c,d}\delta^{b,d}A(t,u,s)
+\delta^{a,d}\delta^{b,c}A(u,t,s)\,,
\end{equation}
where $s,t,u$ are the usual Mandelstam variables.
 The isospin amplitudes $T^{I}(s,t)$ $(I=0,1,2)$ are
\ba
T^{0}(s,t)&=&3A(s,t,u)+A(t,u,s)+A(u,s,t)\,,
\nonumber\\
T^{1}(s,t)&=&A(s,t,u)-A(u,s,t)\,, \quad\quad
T^{2}(s,t)=A(t,u,s)+A(u,s,t)\,,
\ea
and are expanded in partial waves 
\ba
T^{I}(s,t)&=&32\pi\sum_{
  \ell=0}^{+\infty}(2\ell+1)P_{\ell}(\cos{\theta})t_{\ell}^{I}(s),
\ea
where $t$ and $u$ have been written as $t=-\frac{1}{2}(s-4m^{2}_{\pi})(1-\cos{\theta})$,
 $u=-\frac{1}{2}(s-4m^{2}_{\pi})(1+\cos{\theta})$.
Near threshold the $t^I_{\ell}$ are further expanded in terms of the threshold parameters
\begin{eqnarray}
t_{\ell}^{I}(s)=q^{2\ell}( a_{\ell}^{I}+ b_{\ell}^{I}q^{2}+\mathcal{O}(q^{4}))\quad
 q^{2}=\frac{1}{4}(s-4m^{2}_{\pi}) ,
\end{eqnarray}
where $a_{\ell}^{I}, b_{ \ell}^{I}\dots$ are the 
scattering lengths, slopes,$\dots$.
We studied only those observables where a dependence on the $C_{i}$s shows
up.
Using $s+t+u = 4 m_\pi^2$ we can write the amplitude to order $p^6$ as
\begin{eqnarray}
\label{Apipi}
A(s,t,u)=b_{1}+b_{2}s+b_{3}s^{2}+b_{4}(t-u)^{2}+b_{5}s^{3}+b_{6}s(t-u)^{2}
+\textrm{non polynomial part}
\end{eqnarray}
The tree level Feynman diagrams give polynomial contributions to
$A(s,t,u)$ which must be expressible in terms of $b_1,\dots,b_6$.
As a consequence, we obtain the following five relations among the
scattering lengths: 
 \begin{eqnarray}
       \treeC{3b^{1}_1 + 25a^{2}_2} &=& 10\treeC{a^0_2},\\
 \treeC{5b^2_0- 2b^0_0}  + 9\treeC{2b^1_1
  - 3a^1_1}&=& 3\treeC{5a^2_0 - 2a^0_0},\\
  \treeC{- 5b^2_2 + 2b^0_2} &=& 21\treeC{a^1_3},\\
 20\treeC{b^2_2 - b^0_2- a^2_2+a^{0}_2} &=&\treeC{3a^1_1+b^2_0},\\
- 10\treeC{b^2_0 - 18b^0_2+ 18a^0_2}& = &\treeC{2b^0_0 +18a^1_1},
\end{eqnarray}
where the symbol $\treeC{\dots }$ indicates
that these relations are valid for the parts
depending on the $C_i$s only. In fact, since these relations hold for every
contribution to the polynomial part, they are valid at NLO too and both for
$n_f=2$, $3$. 
Therefore they do not get contributions from the $L_i$s at NLO, but
only at NNLO thanks to the non polynomial part of Eq.~(\ref{Apipi}).

\subsection{$\pi K$ scattering}

The $\pi K$ scattering amplitude has amplitudes $T^I(s,t,u)$
in the isospin channels $I= 1/2,3/2$.
 As for $\pi \pi$ scattering, it is possible to define scattering lengths
 $a_{\ell}^{I}$, $b_{\ell}^{I}$. So we introduce the partial wave expansion of
 the isospin amplitudes
\begin{eqnarray}
T^{I}(s,t,u)&=&16\pi
\sum_{\ell=0}^{+\infty}(2\ell+1)P_{\ell}(\cos{\theta})t_{\ell}^{I}(s),
\end{eqnarray}
and we expand the $t_{\ell}^{I}(s)$ near threshold:
\begin{eqnarray}
t_{\ell}^{I}(s)&=&\frac{1}{2}\sqrt{s}q_{\pi K}^{2\ell}
\left(a_{\ell}^{I}+b^{I}_{\ell}q^{2}_{\pi K}+\mathcal{O}(q^{4}_{\pi K})\right),
\qquad
q_{\pi K}^{2}=\frac{s}{4}\left(1-\frac{(m_{K}+m_{\pi})^{2}}{s}\right)
          \left(1-\frac{(m_{K}-m_{\pi})^{2}}{s}\right)\,,
\nonumber
\end{eqnarray}
and
$t=-2q^{2}_{\pi K}(1-\cos{\theta}),\quad u=-s-t+2m^{2}_{K}+2m^{2}_{\pi}$.
Again we studied only those observables where a
dependence on the $C_{i}$s shows up.
 
It is also customary to introduce the crossing symmetric and antisymmetric
amplitudes $T^{\pm}(s,t,u)$ which can be expanded around
$t=0$, $s=u$ using $\nu= (s-u)/(4m_{K})$ (subthreshold expansion):
\begin{eqnarray}
\label{subthrexp}
T^{+}(s,t,u)&=&\sum_{i,j=0}^{\infty}c_{ij}^{+}t^{i}\nu^{2j}, \qquad
T^{-}(s,t,u)=\sum_{i,j=0}^{\infty}c_{ij}^{-}t^{i}\nu^{2j+1}.
\end{eqnarray}
In $c_{01}^{-}$ and $c_{20}^{-}$ the same combination
$-C_{1}+2C_{3}+2C_{4}$ appears~\cite{Bijnens:2004bu}, thus
\begin{equation}
\label{piKrel}
16m^{4}_{K}\treeC{c_{01}^{-}}=3\treeC{c_{20}^{-}}\,.
\end{equation}
Eq.~(\ref{piKrel}) leads to two relations between the scattering lengths which
hold only in the $p^{6}$ case;
 there is a dependence on $L_{3}$ and
$L_{5}$ from the NLO contribution.

\subsection{$\pi\pi$ and $\pi K$ scattering}

 Considering the $\pi\pi$ and $\pi K$ system
 together we get five more relations due to the identities
\begin{eqnarray}
\treeC{b_{5}}&=&\treeC{c^{+}_{30}}+\frac{3}{4m_{K}}\treeC{c^{-}_{20}}\,,
\qquad
\treeC{b_{6}}=\frac{1}{4m_{K}}\treeC{c^{-}_{20}}
   +\frac{1}{16m_{K}}\treeC{c^{+}_{11}},
\end{eqnarray}
where $c^{-}_{ij}$ ($c^{+}_{ij}$) are  expressed
in units of $m^{2i+2j+1}_{\pi}$($m^{2i+2j}_{\pi}$). These relations and those
in the previous subsection are rather long in terms of the threshold
parameters.

\subsection{$K_{e 4}$}
The decay $K^+(p)\to \pi^+(p_1)\pi^-(p_2) e^+(p_\ell)\nu(p_\nu)$ is given 
by the amplitude~\cite{Bijnens:1994me}
\begin{equation}
      T = \frac{G_F}{\sqrt{2}} V^\star_{us} \bar{u} (p_\nu) \gamma_\mu
      (1-\gamma_5) v(p_\ell) (V^\mu - A^\mu)
      \label{k11}
\end{equation}
where $V^\mu$ and $A^\mu$ are parametrized in terms of four form factors: $F$,
$G$, $H$ and $R$  (but the $R$-form factor is negligible in decays with
an electron in the final state).
Using partial wave expansion and neglecting $d$ wave terms one
obtains~\cite{Amoros:1999mg}:
\begin{eqnarray}
\label{Kl4exp}
F_{s}&=&f_{s}+f'_{s}q^{2}+f''_{s}q^{4}+f'_{e}s_{e}/4m^{2}_{\pi}
+\dots\quad (S\textrm{ wave})\,,
\end{eqnarray}
and similar expressions for the other partial waves and form factors.
Here
$s_{\pi}(s_{e})$ is the invariant mass of dipion (dilepton) system, and
$ q^{2}=s_{\pi}/(4m^{2}_{\pi})-1$.
We found one relation involving $F_s$:
\begin{equation}\label{Kl4rel}
\treeC{f''_{s}}F^{4}_{\pi}\frac{\sqrt{2}F_{\pi}}{m_{K}}=64m^{4}_{\pi}\treeC{c^{+}_{30}}\frac{F^{6}_{\pi}}{m^{6}_{\pi}}
\end{equation}
This translates into a relation between $\pi\pi$, $\pi K$ scattering lengths
and $f''_{s}$.

\subsection{Scalar Form Factors and Masses}

The scalar form factors for the pions and the kaons are defined as
\begin{equation}
F_{ij}^{M_{1}M_{2}}(t)=\langle M_{2}(p)|\bar{q}_{i}q_{j}|M_{1}(q)\rangle,
\end{equation}
 where $t=p-q$, $i,j=u, d, s$ are flavour indices and $M_{i}$ denotes a meson
 state with the indicated momentum.
 Due to isospin symmetry not all of them are independent, therefore we
 consider only 
\begin{eqnarray}
 F_{S}^{\pi} &=& 2F^{\pi^0\pi^0}_{uu}\,\qquad\quad
F_{Ss}^{\pi}=F^{\pi^0\pi^0}_{ss}\,,\qquad
F_{Ss}^{K}=F_{ss}^{K^0K^0},
\nonumber\\
F_{S}^{K} &=& F^{K}_{Su}+F^{K}_{Sd}=F^{K^0K^0}_{uu}+F^{K^0K^0}_{dd}\,,\qquad
F_{S}^{\pi K}=F_{su}^{K^+\pi^0}.
\end{eqnarray}
There are two relations between $F_{S}(t=0)$ and the
ChPT expansion of the masses $M^{2}_{\pi},\,M^{2}_{K}$:
\begin{eqnarray}
\label{scalmassrel}
2B_{0}\treeC{M^{2}_{\pi}}&=&\frac{1}{3}
      \left\{(2m^{2}_{K}-m^{2}_{\pi})\treeC{F^{\pi}_{Ss}(0)}
                  +m^{2}_{\pi}\treeC{F^{\pi}_{S}(0)}\right\}
\nonumber\\
2B_{0}\treeC{M^{2}_{K}}&=&\frac{1}{3}
      \left\{(2m^{2}_{K}-m^{2}_{\pi})\treeC{F^{K}_{Ss}(0)}
               +m^{2}_{\pi}\treeC{F^{K}_{S}(0)}\right\}.
\end{eqnarray}
One could arrive to the same conclusion using the Feynman-Hellmann Theorem
(see e.g.~\cite{Gasser:1984ux} or~\cite{Bijnens:2003xg}) which implies
for $q=u,d,s$ and $M=\pi,K$
\begin{eqnarray}
F^{M}_{Sq}(t=0) &=&
  \langle M|\bar{u}u|M\rangle =
  \frac{\partial m^{2}_{M}}{\partial m_{q}}\,.
\end{eqnarray}
On the other hand the ChPT expansion leads to
\begin{eqnarray}
\treeC{M^{2}_{\pi}}=\sum_{i}C_{i}(m_{q})^{3}=f(m_{u},m_{d},m_{s}),
\end{eqnarray}
that is an homogeneous function of order three.
Thanks to the Euler Theorem, $\treeC{M^{2}_{\pi}}$ can be written in terms
of its derivatives $(f(\mathbf{x})=\frac{1}{3}\sum^{n}_{i=1}\frac{\partial
  f}{\partial x_{i}}x_{i}\quad\mathbf{x}\in \mathbb{R}^{n})$. 
These are exactly the relations in Eq.~(\ref{scalmassrel}).
Something similar holds for the $p^4$ expression but with a factor
$1/2$ instead of $1/3$.

\subsection{Other Relations}

Here we present just a general overview of the other relations found:
\begin{itemize}
\parskip0cm\itemsep0cm
\item Decay constants, Masses and Scalar Form Factors : two more relations
\item Vector Form Factors: no new non trivial relation
\item $\eta \rightarrow 3\pi$: no relations
\item Considering all together Scalar Form Factors, Masses, Decay Constants, $\pi\pi$ scattering and $\pi K$ scattering : one extra (difficult) relation,
essentially the equivalent of the relation in \cite{BT3}.
\item Another relation between $K_{\ell 4}$ form factors ($F_{p}$, $G_{s}$,
  $G'_{s}$, $G''_{s}$), $\pi K$ and $\pi\pi$ coefficients, and scalar form factors.
\end{itemize}

\section{Numerical Analysis}

In this section we present first results of a numerical analysis of the
relations appearing in Eqs.~(\ref{piKrel}) and (\ref{Kl4rel}). For both of
them we evaluated numerically the relevant quantities (i.e. $T^\pm$ for $\pi
K$ scattering and $F_s$ for $K_{\ell 4}$) setting for the $L_i$s the values of
fit 10 and the $C_i$s$=0$. Then we performed a fit to the
expressions in Eqs.~(\ref{subthrexp}) and~(\ref{Kl4exp}) respectively. 
This is the part of the quantities that does not come from the $C_i$
and which needs to be subtracted from the experimental results to test
the relations.
The experimental part is evaluated
from the dispersive results for $\pi K$  scattering~\cite{Buettiker:2003pp}
and experiment for $K_{\ell 4}$~\cite{Batley:2007zz,Pislak:2001bf}.
We found that the relations are not well satisfied. The reason
for these discrepancies is still under investigation.

\subsection{$\pi K$ scattering}

\begin{figure}[h]
\vskip-0.3cm
\begin{minipage}{0.47\textwidth}
\includegraphics[angle=270, width=0.99\textwidth]{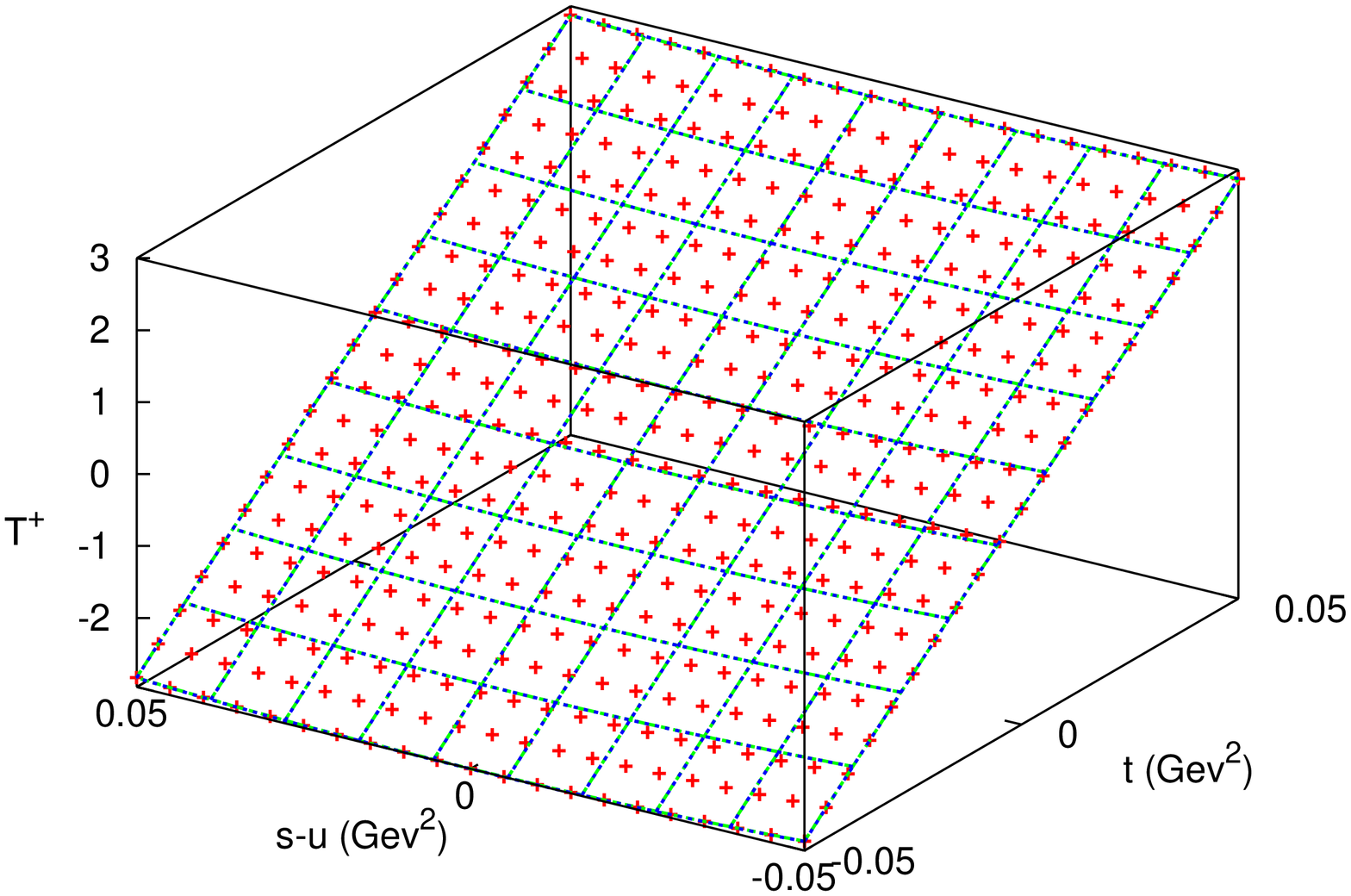}
\caption{\label{piKTplus}$T^+$ as a function of $t$ and $s-u$. 
  Red points are numerics
  generated with $L_i$s$=$fit10 and $C_i$s$=0$. 
  Fitting is with $\sum_{i}\sum_{j}c^{+}_{ij}t^{i}\nu^{2j+1}$. 
  Blue surface: $i+2j\le 5$. Green surface: $i+2j\le 4$}
\end{minipage}
~~~
\begin{minipage}{0.47\textwidth}
\includegraphics[angle=270, width=0.99\textwidth]{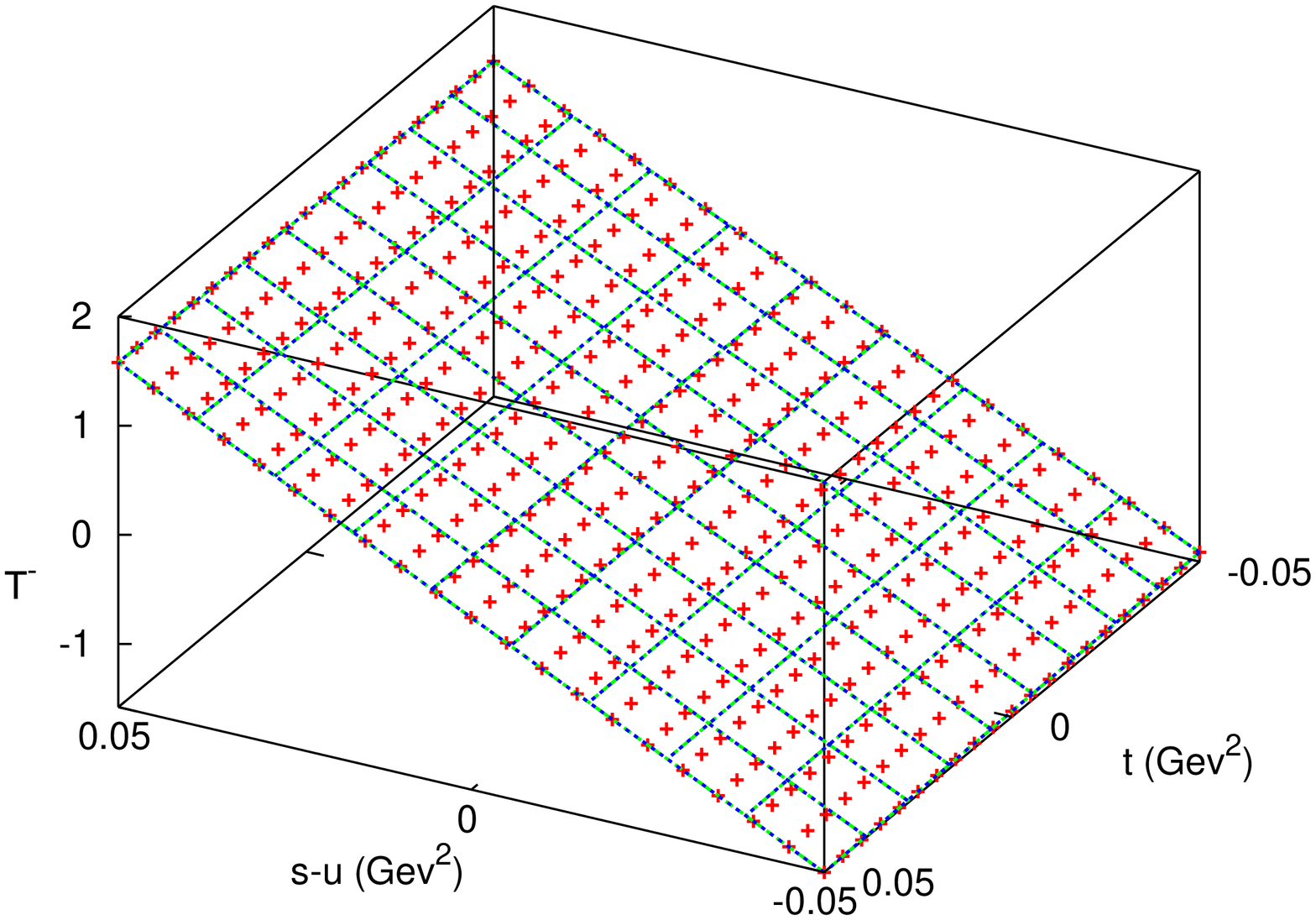}
\caption{\label{piKTminus}$T^-$ as a function of $t$ and $s-u$. 
  Red points are numerics
  generated with $L_i$s$=$fit10 and $C_i$s$=0$. 
  Fitting is with $\sum_{i}\sum_{j}c^{+}_{ij}t^{i}\nu^{2j+1}$. 
  Blue surface: $i+2j\le 5$. Green surface: $i+2j\le 4$}
\end{minipage}
\end{figure}

The fit of the subthreshold expansion (\ref{subthrexp}) to the ChPT NNLO
result is shown in Figs.~\ref{piKTplus} and~\ref{piKTminus}
(notice there are three surfaces in each plot)
and the resulting to be subtracted threshold parameters
are shown in Tabs.~\ref{tab_piKTplus} and~\ref{tab_piKTminus}.
\begin{table}
\begin{multicols}{2}
\begin{tabular}{|c|c|c|c|}
\hline
\multicolumn{4}{|c|}{Subthreshold parameters ($C_i$s$=0$, $L_i$s$=$fit10)}\\
\hline
$10c^{+}_{00}$ & $-0.709$ & $c^{+}_{10}$ & $1.101$ 
\\ \hline $10^{2}c^{+}_{20}$ & $-0.485$ 
& 
$c^{+}_{01}$ & $3.467$  
\\ \hline
$10^{2}c^{+}_{30}$ & $0.186$ 
&
$c^{+}_{11}$ & $-0.131$ 
\\ 
\hline
$10^{3}c^{+}_{40}$ & $0.250$ 
&
$10^{2}c^{+}_{21}$ & $0.824$ 
 \\ \hline 
\end{tabular}
\caption{ Values of $c^+_{ij}$ (in unit of  $m^{2(i+j)}_\pi$) from fit with $i+2j\le 5$ (see Figure~\protect\ref{piKTplus})}
\label{tab_piKTplus}

\begin{tabular}{|c|c|c|c|}
\hline
\multicolumn{4}{|c|}{ Subthreshold parameters ( $C_i$s$=0$, $L_i$s$=$fit10)}\\
\hline
$c^{-}_{00}$ & $8.398$ & $10c^{-}_{10}$ & $0.959$ 
\\ \hline $10^{2}c^{-}_{20}$ & $0.791$ 
& 
$c^{-}_{01}$ & $0.426$  
\\ \hline
$10^{4}c^{-}_{30}$ & $-1.04$ 
&
$10^{2}c^{-}_{11}$ & $-6.04$ 
\\ 
\hline
\end{tabular}
\caption{Values of $c^-_{ij}$ (in units of $m^{2i+2j+1}_\pi$) from fit
  with $i+2j\le 5$ (see Figure~\protect\ref{piKTminus} ).}
\label{tab_piKTminus}
\end{multicols}
\end{table}
Using the results of Tab.~\ref{tab_piKTminus} we get for the to be
subtracted part of the relation (\ref{piKrel}):
\begin{equation}
\label{piKone}
\frac{F^{6}_{\pi}}{m^{3}_{\pi}} (1.582)
     \iff
\frac{F^{6}_{\pi}}{m^{3}_{\pi}}( 1.278)\,.
\end{equation}
The dispersive analysis~\cite{Buettiker:2003pp} gives the experimental
results for both sides of (\ref{piKrel})
\begin{equation}\label{piKtwo}
\frac{F^{6}_{\pi}}{m^{3}_{\pi}} (1.70 \pm 0.02 )
         \iff
\frac{F^{6}_{\pi}}{m^{3}_{\pi}}( 1.9 \pm 0.18)
\end{equation}
The difference (\ref{piKtwo})$-$(\ref{piKone}) is what should satisfy
(\ref{piKrel}):
\begin{equation}
\label{piKdiscr}
0.12 \pm 0.02 \stackrel{\textstyle ?}{=} 0.6 \pm 0.18
\end{equation}
As you see in (\ref{piKdiscr}) the right and the left hand side are not
in agreement.
Probably it is the same discrepancy
found between ChPT~\cite{Bijnens:2004bu}, $c^{-}_{20}=0.013$, and
dispersive results~\cite{Buettiker:2003pp} $c^{-}_{20}=0.0085 \pm 0.0001$
noticed before \cite{Bijnens:2004bu}.
This is related to the conflicting $C_i$ determinations
in~\cite{Kampf:2006bn}.

\subsection{$K_{l4}$}

We now do the same analysis for $F_s$.
As shown in Figure~\ref{kl4fig} we performed two fits with different degree
polynomials. The higher degree fits better the dependence on $s_\ell$ even
thought in the region probed experimentally in~\cite{Batley:2007zz}
$(\frac{s_\ell}{4m^2_\pi}\leq 0.4 \textrm{ and } q^2\leq 1)$ the lower degree
polynomial fits well. $f_{s}$ and $f_{e}$ turns out to be in agreement between
the two fits. We quote here the results for the blue fit. The uncertainties are
a measure on how much the two fits differ:
\begin{figure}[h]
\begin{center}
\includegraphics[angle=270,width=6cm]{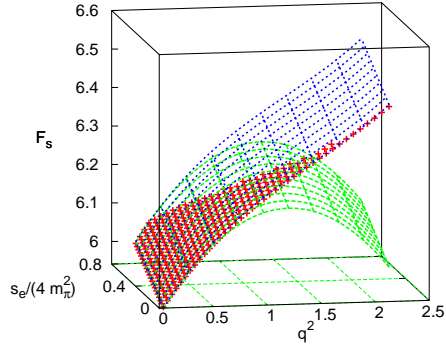}
\caption{\label{kl4fig}$F_s$ as a function of $\frac{s_\ell}{4m_\pi}$ and $q^2$. Red point are numerics
  generated with $L_i=$fit10 and $C_i=0$. Green surface: fit with
  $f_{s}(1+\frac{f'_{s}}{f_{s}}q^{2}+\frac{f''_{s}}{f_{s}}q^{4}+\frac{f_{e}}{f_{s}}\frac{s_{e}}{4m^{2}_{\pi}})$. Blue
  surface: fit with
  $f_{s}(1+\frac{f'_{s}}{f_{s}}q^{2}+\frac{f''_{s}}{f_{s}}q^{4}+\frac{f'''_{s}}{f_{s}}q^{6}+\frac{f_{e}}{f_{s}}\frac{s_{e}}{4m^{2}_{\pi}})$
}
\end{center}
\end{figure}
\begin{eqnarray}
\label{Kl4fit}
f_{s}=5.924 \pm 0.002 \quad \frac{f'_{s}}{f_{s}}=0.075 \pm 0.005 
\quad \frac{f''_{s}}{f_{s}}=-0.03 \pm 0.009 \quad \frac{f'_{e}}{f_{s}}=0.038 \pm 0.002
\end{eqnarray}
Using the value (\ref{Kl4fit}) for $f''_s$, Tab.~\ref{tab_piKTplus} for
$c^+_{31}$ we get for the to be subtracted part for the relation (\ref{Kl4rel})
\begin{equation} 
\label{kl4one}
-0.049 \pm 0.002 \iff 0.0521 \pm 0.0006\,.
\end{equation}
The experimental results
of~\cite{Batley:2007zz}  (value for $f_{s}=5.77$ from~\cite{Pislak:2001bf})
and ~\cite{Buettiker:2003pp} give the experimental part:
\begin{equation}
\label{kl4two}
-0.14 \pm 0.04 \iff 0.09 \pm 0.02\,.
\end{equation}
The difference (\ref{kl4two})$-$(\ref{kl4one}) is:
\begin{eqnarray}
-0.08 \pm 0.04 \stackrel{\textstyle ?}{=} 0.04 \pm 0.02\,.
\end{eqnarray}
Again a discrepancy shows up: 
now the two sides of the relation~(\ref{Kl4rel}) have opposite sign.

\section{Conclusions}

We have presented here the first results of search for relations
at NNLO in ChPT that are independent of the order $p^6$ LECs.
We found several previously unknown relations and have presented
preliminary numerical results for two of them. 

\section*{Acknowledgments}

IJ gratefully acknowledges an Early Stage Researcher position supported by the
EU-RTN Programme, Contract No. MRTN--CT-2006-035482, (Flavianet)
This work is supported in part by the European Commission RTN network,
Contract MRTN-CT-2006-035482  (FLAVIAnet), 
European Community-Research Infrastructure
Integrating Activity
``Study of Strongly Interacting Matter'' (HadronPhysics2, Grant Agreement
n. 227431)
and the Swedish Research Council. We thank the organizers for a very pleasant
meeting.

\end{document}